\documentclass[conference]{IEEEtran}
\IEEEoverridecommandlockouts
\usepackage{cite}
\usepackage{amsmath,amssymb,amsfonts}
\usepackage{algorithmic}
\usepackage{graphicx}
\usepackage{textcomp}
\usepackage{xcolor}
\usepackage{hyperref}

\newcommand{\Sref}[1]{\S\ref{#1}}
\newcommand{\fref}[1]{figure~\ref{#1}}

\newcommand{\Eref}[1]{Equation~\ref{#1}}

\usepackage{booktabs,tabularx}
\def\BibTeX{{\rm B\kern-.05em{\sc i\kern-.025em b}\kern-.08em
    T\kern-.1667em\lower.7ex\hbox{E}\kern-.125emX}}
\begin{document}

\title{Affective Polarization in Online Climate Change Discourse on Twitter\\ 
\thanks{This work was supported in part by the Knight Foundation and the Office of Naval Research grants N000141812106 and N000141812108. Additional support was provided by the Center for Computational Analysis of Social and Organizational Systems (CASOS), the Center for Informed Democracy and Social Cybersecurity (IDeaS), and the Department of Engineering and Public Policy of Carnegie Mellon University. The views and conclusions contained in this document are those of the authors and should not be interpreted as representing the official policies, either expressed or implied, of the Knight Foundation, Office of Naval Research or the U.S. government.}
}

\author{\IEEEauthorblockN{Aman Tyagi}
\IEEEauthorblockA{\textit{CASOS, Engineering and Public Policy} \\
\textit{Carnegie Mellon University}\\
PIttsburgh PA, USA \\
amantyagi@cmu.edu}
\and
\IEEEauthorblockN{Joshua Uyheng}
\IEEEauthorblockA{\textit{CASOS, Institute for Software Research} \\
\textit{Carnegie Mellon University}\\
Pittsburgh PA, USA \\
juyheng@cs.cmu.edu}
\and
\IEEEauthorblockN{Kathleen M. Carley}
\IEEEauthorblockA{\textit{CASOS, Institute for Software Research}\\
\textit{Carnegie Mellon University}\\
Pittsburgh PA, USA \\
kathleen.carley@cs.cmu.edu}
}

\maketitle

\begin{abstract}
Online social media has become an important platform to organize around different socio-cultural and political topics. An extensive scholarship has discussed how people are divided into echo-chamber-like groups. However, there is a lack of work related to quantifying hostile communication or \textit{affective polarization} between two competing groups. This paper proposes a systematic, network-based methodology for examining affective polarization in online conversations. Further, we apply our framework to 100 weeks of Twitter discourse about climate change. We find that deniers of climate change (Disbelievers) are more hostile towards people who believe (Believers) in the anthropogenic cause of climate change than vice versa. Moreover, Disbelievers use more words and hashtags related to natural disasters during more hostile weeks as compared to Believers. These findings bear implications for studying affective polarization in online discourse, especially concerning the subject of climate change. Lastly, we discuss our findings in the context of increasingly important climate change communication research.
\end{abstract}


\section{Introduction}

Online social networks represent a powerful space for public discourse. Through large-scale, interconnected platforms like social media, diverse communities may potentially participate in open exchanges of views and information about a vast range of issues. However, research has increasingly demonstrated the dangers of \textit{polarization} in online communication \cite{barbera2015tweeting,garimella2018political,tyagi2020computational}. Attributed to various psychological, social, and technological factors, intergroup communication on cyberspace has displayed tendencies to feature pathological dynamics especially concerning contentious issues \cite{brady2017emotion,geschke2019triple}. 

Polarization on social media could be broadly divided into different categories. Opposed groups may communicate in a highly balkanized fashion, such that members of an in-group are only minimally exposed to out-group members and their beliefs \cite{karlsen2017echo,matakos2017measuring}. This phenomenon has been termed \textit{interactional polarization}. Polarization can also pertain to highly negative sentiments toward out-groups in the form of \textit{affective polarization} \cite{anderson2017social,druckman2019we}. Social scientific research examines how these phenomena are interconnected across a variety of contexts, such that online groups that disagree on a given topic are also more likely to be hostile toward each other \cite{yarchi2020political}. In this paper, we focus on quantifying affective polarization between two groups with opposing beliefs using Twitter discourse on a significant social issue.

One significant issue which has received heated attention in online public discourse is climate change \cite{dunlap2016political,fisher2013does,mccright2011politicization,tyagi2020brims}. We focus on those who cognitively accept anthropogenic causes of climate change (\textit{Believers}) and those who reject the same (\textit{Disbelievers}). Previous work demonstrates not only sharp divergences in climate change beliefs but also the emergence of communities insulated from the opposed group \cite{hamilton2015tracking,milfont2017public,tyagi2020brims}. In other words, online discussions about climate change are \textit{interactionally polarized}, implying the persistence of echo chambers between Believers and Disbelievers \cite{jang2015polarized,van2020online,williams2015network}.   

Much less work, however, engages the question of \textit{affective polarization} in online climate change discourse. A crucial limitation in prior work lies in the methodological options available to past researchers. Relying consistently on manually annotated corpora and datasets of limited size, existing scholarship has faced barriers to measuring the emotional component of climate change discussions in a generalizable fashion \cite{anderson2017social,jang2015polarized,van2020online}. Drawing on recent advances in computational stance detection, targeted sentiment analysis, and network science measures, we present an integrated methodological pipeline for addressing this gap in the literature. 

More specifically, we show how computational methods may be leveraged to generate (a) automated stance labels for climate change Believers and Disbelievers, (b) individual measurements of the interaction valence between in-group and out-group members, and (c) broader assessments of group-level affective polarization. We demonstrate the utility of our framework by applying our methodology to a large-scale dataset of 100 weeks of online climate change discussion on Twitter. Furthermore, we link our findings to natural disasters words to explain important climate change belief constructs.

In sum, this work proposes to answer the following research questions:
\begin{enumerate}
    \item How can affective polarization be computationally measured on a large-scale, long-term corpus of online climate change discussions?
    \item Do climate change Believers or Disbelievers feature greater levels of affective polarization in online public discourse?
    \item What is the relationship of affective polarization between the two groups with use of natural disaster related words\footnote{We provide the list of natural disaster related words used in our analysis in \Sref{sec:appendix} }?
\end{enumerate}

The subsequent sections of this paper are organized as follows. First, we provide an overview of related work in this area, illustrating computational analysis of polarization in general terms and then in the case of climate change specifically. We zero in on the dearth of principled empirical work on affective polarization specifically in relation to online climate change discourse. Second, we present our proposed methodological pipeline which integrates machine learning models and network science techniques to facilitate a novel and effective framework for assessing affective polarization. Third, we share our findings on our large-scale, long-term Twitter dataset. Last, we discuss implications for understanding the state of climate change discourse on digital platforms as well as related empirical investigation of affective polarization on online social networks.

\section{Related Work}

\subsection{Computational analysis of polarization}

Recognizing the ubiquity of online conflicts, rigorous scholarship in the computational and social sciences has tackled the problem of polarization. More traditional approaches in offline settings have relied on survey measures to empirically assess divergence in beliefs between groups \cite{banks2020polarizedfeeds,druckman2019we}. But with burgeoning developments in computational methods - especially with respect to natural language processing and machine learning - automated methods have also arisen to leverage the vast digital traces linked to online activity \cite{huang2018aspect,kumar2020social}.

General approaches infer individual attitudes from user information, such as the texts associated with an account on social media (e.g., Facebook comments, tweets). Group membership as well as group communication are similarly incorporated into analyses of polarization, by examining the beliefs of individuals in conjunction with their traceable patterns of digital interaction with other individuals. Given various conceptualizations of polarization, different frameworks have been developed to quantify pathological patterns of communication across groups holding similar or opposed stances on a given issue \cite{darwish2019quantifying,demszky-etal-2019-analyzing,morales2015measuring,Weber2013}.

Social network analysis has gained much methodological traction in this regard. Representing online conversations as graphical structures, numerous approaches measure polarization as a function of homophily in local community structures \cite{matakos2017measuring,stewart2018examining}. In other words, the extent to which those holding similar views are more likely to interact with each other - in contrast to those with whom they disagree - allows an intuitive and principled measure of polarization. For example, random walk scores quantify the probability of a random walk starting from a node belonging to a given stance group ending up in a node belonging to the same or a different stance group \cite{garimella2018political,garimella2017reducing,garimella2018quantifying}.

More recent scholarship, however, emphasizes the importance of examining not just pathologically isolated communication, or interactional polarization; but also pathologically hostile communication, or affective polarization. Burgeoning evidence suggests that the problem of echo chambers represents a significant, yet incomplete, picture of polarization in online social networks \cite{barbera2015tweeting}. People holding opposed views, in fact, do interact with each other - but this does not necessarily mitigate polarization \cite{karlsen2017echo}. Instead, research finds that these intergroup exposures trigger further incivility and toxicity \cite{banks2020polarizedfeeds}. Hence, reliable measures for affective polarization are needed, although the computational literature in this area remains in its nascent stages \cite{yarchi2020political}.

\subsection{Climate change and polarization}

In the specific case of climate change discourse, analysis of polarization has also represented a major research topic. Numerous studies link polarized beliefs about climate change to partisan divides, with more conservative individuals less likely to cognitively accept anthropogenic climate change than liberals \cite{dunlap2016political,hamilton2015tracking}. Past work specifically demonstrates that although higher levels of education and information access may increase the likelihood of climate change belief, these effects remain much lower among conservatives \cite{hamilton2015tracking,mccright2011politicization}. Such effects have been explained from the lens of elite signalling - whereby followers emulate the beliefs of their preferred political leaders - uneven exposure to information based on partisan media, as well as a generalized dislike for the members of the opposed ideological group \cite{bolsen2018us,carmichael2017elite,van2018psychological}.

However, with time, scholars have also noted general trends toward increasing climate change beliefs overall \cite{milfont2017public}. Even if these do not necessarily translate into concrete support for policy instruments to address climate issues \cite{fisher2013does}, the long-term instability of climate change skepticism points to valuable ways forward for science communication \cite{jenkins2020partisan}. Collectively, these finding suggest the importance of accounting for the psychological processes surrounding climate change belief and disbelief, going beyond the transmission of information \cite{kahan2015geoengineering}. 

These issues take on specific forms in cyberspace, where information flows are inextricably entangled with community dynamics. Studies employing social network analysis have uncovered robust evidence that online climate change discussions tend to exhibit echo chamber-like homophilic interactions \cite{bloomfield2019circulation,williams2015network,tyagi2020brims}.   Qualitative analysis further showed that in rare instances of intergroup communication, more negative frames tended to prevail, featuring dismissal of climate change information as hoaxes, derailment of conversations to heated issues of identity, as well as overall higher levels of sarcasm and incivility \cite{anderson2017social,jang2015polarized,van2020online}. Notwithstanding the valuable idiographic insights derived from these studies, their sampling strategies have tended to rely on a minuscule fraction of the larger conversation to facilitate in-depth content analysis. Hence, larger-scale and more generalizable findings on the affective dynamics of online climate change discourse are notably lacking in the literature.

\subsection{Contributions of this work}

Motivated by the foregoing insights, this work seeks to contribute to the literature by offering a methodological pipeline for examining affective polarization. As the succeeding sections demonstrate, our framework combines machine learning and network science methods in a novel, scalable, and generalizable fashion for ready application in a variety of contentious issues. This overcomes many of the methodological barriers present in prior work, including their common reliance on expensive survey or experimental measures, or manually annotated datasets in the context of social media research on climate change discourse \cite{carmichael2017elite,hamilton2015tracking,mccright2011politicization,milfont2017public,williams2015network}.

From a theoretical standpoint, we additionally contribute a nuanced operationalization of affective polarization as located on a group level. We unpack how group-level metrics valuably produce asymmetrical views of hostile behavior, thereby facilitating more fine-grained analysis of how different stance groups engage in varied levels of affectively polarized interactions. This conceptually aligns with the asymmetry of psychological factors characterizing climate change Believers and Disbelievers, especially over time \cite{dunlap2016political,jenkins2020partisan,van2018psychological}.

Finally, on an empirical level, our work also extends prevailing scholarship on polarized climate change discourse. While established findings paint a picture of consistent echo chambers between climate change Believers and Disbelievers, we provide evidence for the flipside of these dynamics. We specifically quantify, over a larger-scale and longer-term dataset than previously examined in prior work, the extent to which intergroup interactions systematically feature hostility. This may inform possible data-driven interventions for policymaking beyond more prevalent frames of intergroup contact and science communication \cite{garimella2017reducing,kahan2015geoengineering}.
\section{Data and Methods}

\subsection{Data collection}
\label{sec:datacollection}

We collected realtime tweets using Twitter's standard API\footnote{\url{https://developer.Twitter.com/en/docs/tweets/search/overview/standard}} with keywords ``Climate Change'', ``\#ActOnClimate'', ``\#ClimateChange''. Our dataset was collected between August 26th, 2017 to September 14th, 2019. Due to server errors, the collection was paused from April 7th, 2018 to May 21st, 2018, and again from May 12th, 2019 to May 16th, 2019. We ignore these periods from our analysis. We de-duplicated the collected tweets to remove any duplicate tweets collected more than once. Overall, we collected 38M unique tweets and retweets from 7M unique users. For our analysis, we aggregate tweets from each user for seven day period (1 week) to get a total of 100 weeks. 

\subsection{Stance labels}

We use a state-of-the-art stance mining method \cite{kumar2020social} to label each user as a climate change Disbeliever or Believer. We use a weak supervision based machine learning model to label the users in our dataset.  The model uses a co-training approach with label propagation and text-classification. The model requires a set of seed hashtags essentially being used by Believers and Disbelievers. The model then labels seed users based on the hashtags used at the end of the tweet. Using the seed users, the model trains a text classifier and uses a combined user-retweet and user-hashtag network to propagate labels. In an iterative process, the model then labels users who are assigned a label by both methods with high confidence\footnote{ We use the parameter values as defined in \cite{kumar2020social} as \{$k = 5000$, $p = 5000$, $\theta^{I} = 0.1$, $\theta^{U} = 0.0$, $\theta^{T} = 0.7$\}.}.

We set \textit{ClimateChangeIsReal} and \textit{SavetheEarth} as Believers seed hashtags and \textit{ClimateHoax} and \textit{Qanon} as Disbelievers seed hashtags. These hashtags have been shown to be used mostly by the respective groups\cite{tyagi2020brims}. Out of the total 7M users, the algorithm labels 3.9M as Believers and 3.1M as Disbelievers. We randomly sampled 500 users from each group to manually validate the results. We label a user as Disbeliever if we find any Tweet akin to someone who does not believe in climate change or anthropogenic cause of climate change. Otherwise, we label the user as Believer. We observe that the average precision from manual validation of 1000 users is 81.8\%. 

\subsection{Affective polarization metrics}

We measure affective polarization in this work by combining outputs from an aspect-level sentiment model, a classic network science measure known as the E/I index \cite{krackhardt1988informal} and Earth Mover's Distance (EMD) \cite{hitchcock1941distribution}. Outputs are combined in the five steps which follow to produce dynamic group-level measurements of affective polarization.

\subsubsection{Aspect-level sentiment} 

Aspect-level sentiment refers to the emotional valence of a given utterance toward one of the concepts it mentions \cite{huang2018aspect}. Sentiments toward specific entities are vital to consider in polarized discussions such as those we consider here. For instance, climate change Disbelievers might express negative feelings toward notions of greenhouse gases, while in agreement with a fellow Disbelievers with whom they are interacting. 

We utilize Netmapper to extract entities from each tweet, and predict the aspect-level sentiment of each tweet toward each entity \cite{carleyora2018,uyheng2019interoperable}. Netmapper relies on a multilingual lexicon of positively and negatively valenced words to calculate sentiment values. Aspect-level sentiment relies on a heuristic of a sliding window over words in the sentence. More specifically, word-level sentiment is computed based on the average of known valences for surrounding words. 

For the purposes of this work, each tweet by a certain agent $i$ which mentions or replies to agent $j$ is assigned an aspect-level sentiment score from $-1$ (very negative) to $+1$ (very positive) directed toward the concept ``@[agent $j$'s Twitter handle]'' \cite{uyheng2019interoperable}. This allowed us to compute affective dimensions to the communication between groups of the same or opposed stance groups.

\subsubsection{Affective networks} 

Given the aspect-level sentiment scores, we construct two affective networks representing the climate change conversations on a per-week basis. Let $G^{+} = (V, E^{+})$ denote a positive interaction network where the set of vertices $V$ contains all Twitter accounts in our dataset and the set of directed edges $E^{+}$ contains all positive-valenced mentions and replies between agents in $V$. Similarly, let $G^{-} = (V, E^{-})$ denote a negative interaction network over the same set of agents $V$ and the set of directed edges $E^{-}$ representing their negative-valenced mentions and replies. 

In both cases, $E^{+}$ and $E^{-}$ denote weighted edges. We obtain their weights as follows. Let $S_{ij}$ denote the set of all aspect-level sentiments in tweets by agent $i$ toward agent $j$, where $i, j \in V$. Then the weight $w_{ij}^{+}$ of edge $e_{ij}^{+} \in E^{+}$ from $i$ to $j$ is given by $\sum_{x \in S_{ij}}\min{(0, x)}$. Conversely, the weight $w_{ij}^{-}$ of edge $e_{ij}^{-} \in E^{-}$ from $i$ to $j$ is given by $\sum_{x \in S_{ij}} \min{(0, -x)}$.

\subsubsection{E/I indices} 

We assess group-level differences in positive and negative interactions using Krackhardt's E/I index \cite{krackhardt1988informal}. For a given affective network, the E/I index intuitively captures the extent to which each stance group $k$ engages in correspondingly valenced interactions with members of the out-group relative to their in-group. Hence, for instance, high values of the E/I index for the negative interaction network would indicate that the given stance group interacts in a more negative way to their opponents relative to those who share their beliefs. 

To compute the E/I indices, let $V_k \subseteq V$ denote the set of agents belonging to stance $k$ and $V_{k'}$ those who do not hold stance $k$. The E/I index of stance group $k$ on the positive interaction network is therefore computed as follows:

\begin{equation}
    P_k^{+} = \dfrac{E_k^{+} - I_k^{+}}{E_k^{+} + I_k^{+}}
\end{equation}
where $E_k^{+} = \sum_{i \in V_k, j \in V_{k'}} w_{ij}^{+}$ and $I_k^{+} = \sum_{i,j \in V_k} w_{ij}^{+}$. On the other hand, the E/I index of stance group $k$ on the negative interaction network is similarly computed thus:

\begin{equation}
    P_k^{-} = \dfrac{E_k^{-} - I_k^{-}}{E_k^{-} + I_k^{-}}
\end{equation}
where $E_k^{-} = \sum_{i \in V_k, j \in V_{k'}} w_{ij}^{-}$ and $I_k^{-} = \sum_{i,j \in V_k} w_{ij}^{-}$. Given the construction of $P_k^{+}$ and $P_k^{-}$, we note that both values are bounded between $-1$ and $+1$.

\subsubsection{Polarization valence}
\label{sec:valence}

We find whether the interactions have negative valence or positive valence by defining polarization $P_k$ by taking a  difference of the two E/I indices as expressed below:  

\begin{equation}
    P_k = P_k^{-} - P_k^{+}.
\end{equation}

In this work, we operationalize our view of affective polarization in terms of high E/I indices on the negative interaction network, and low values on the positive interaction network.  $P_k$ thus captures this intuition by assigning positive values for groups that display disproportionately hostile or negative interactions toward the out-group relative to their in-group. Values close to $0$, on the other hand, indicate relatively even levels of positive and negative interactions for in-group and out-group members. Finally, negative values indicate that those holding stance $k$ are more negative to their in-group but positive to their out-group. 


\subsubsection{Polarization magnitude} 

To find the magnitude of affective polarization we use Earth Mover's Distance (EMD) on the distribution of weighted edges for outgroup and ingroup interactions. This is similar to computing first Wasserstein distance between two 1D distributions\cite{ramdas2017wasserstein}. Similar to affective networks, we define $G = (V, E)$ as interaction network where the set of vertices $V$ contains all Twitter accounts in our dataset and the set of directed edges $E$ contains all valenced (positive or negative) mentions and replies between agents in $V$. In this case, we do not separate negative and positive valence graphs and treat weight $w_{ij}$ of edge $e_{ij} \in E$ from $i$ to $j$ as given by $\sum_{x \in S_{ij}}x$. Let $u_{k}$ be distribution of $w_{ij}$, where $i \in V_k, j \in V_{k'}$ and let $v_{k}$ be distribution of $w_{ij}$, where $i \in V_k, j \in V_{k}$. For a group holding stance $k$, we define our novel affective polarization metric as:  

\begin{equation}
l_{k} = \left\{
  \begin{array}{lr}
    -\int_{-\infty}^{+\infty}|U_{k} - V_{k}| & : P_{k} < 0\\
    \int_{-\infty}^{+\infty}|U_{k} - V_{k}| & : P_{k} \ge 0
  \end{array}
\right.
\label{eq:metric}
\end{equation}

where $U_{k}$ and $V_{k}$ are the respective CDFs of $u_{k}$ and  $v_{k}$. 
EMD is proportional to the minimum amount of work required to covert one distribution to another \footnote{\url{http://infolab.stanford.edu/pub/cstr/reports/cs/tr/99/1620/CS-TR-99-1620.ch4.pdf}}. We use $P_{k}$ to assign positive or negative valence to the EMD. Although there are other techniques to find the difference in distribution such as KS-Test \cite{massey1951kolmogorov}. However, during our experiments, we found that EMD is able to capture more nuanced differences in distributions. More likely because the EMD can capture differences in heavy-tailed distributions better and it does not make any parametric assumptions \cite{ramdas2017wasserstein}. 

Our novel affective polarization metric $l_{k}$ is positive when $P_{k} > 0$. As noted in \Sref{sec:valence}, a positive value would mean more hostility or negative sentiment in intergroup communication compared to intragroup communication. On the other hand, a negative value of $l_{k}$  is when $P_{k} < 0$, meaning more positive sentiment in intergroup communication compared to intragroup communication.
\section{Results}

Using the metric defined in \Eref{eq:metric}, in this section, we first explore how affective polarization between Believers and Disbelievers is changing over the 100 weeks. Then we explore how hostile periods are related to natural disaster-related words.

\begin{figure*}[ht]
    \centering
    \includegraphics[width=0.85\linewidth]{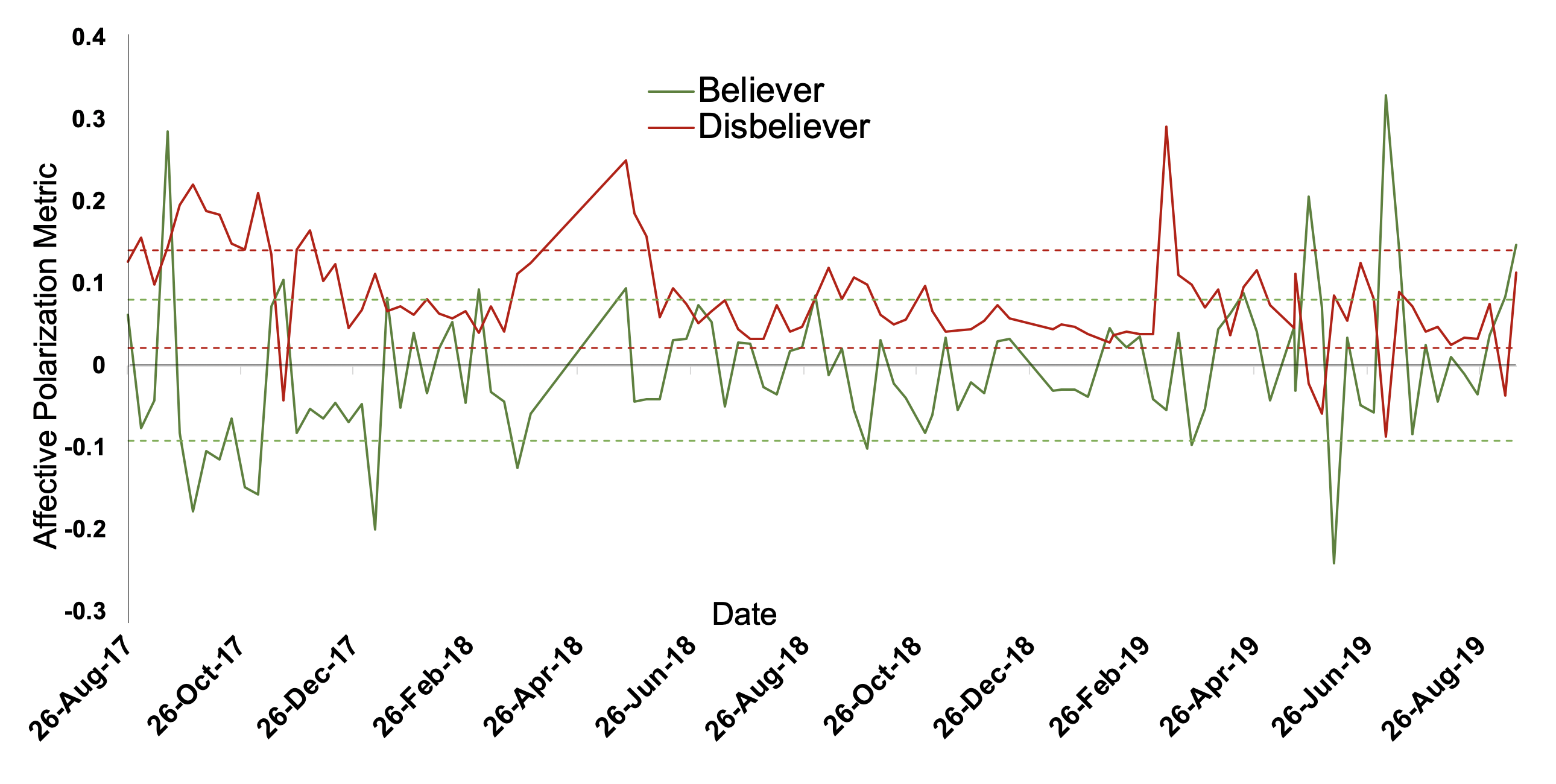}
    \caption{Affective polarization metric ($l_{k}$) for Believers and Disbelievers of climate change. Higher positive values denote more hostility towards the other group. The dotted lines represent mean ±1 standard deviation, which for Believers is -0.091 and 0.080 and disbelievers is -0.117 and 0.106. The analysis was done on data collected from 26th August 2017 to 14th September 2019 as described in \Sref{sec:datacollection}.}
    \label{fig:timeline}
\end{figure*}

We first look at how the affective polarization metric is changing over time in \fref{fig:timeline}. Overall, our analysis found that climate change Disbelievers tended to exhibit high levels of hostility toward climate change Believers. This finding was relatively consistent throughout the 100-week period under observation, as the time series for climate change Disbelievers only very rarely goes below the threshold of 0, which indicates similarly valenced interactions toward in-group and out-group members. Some weeks displayed exceptionally high levels of hostility toward climate change Believers, greater than one standard deviation from the mean. The standard deviation of $l_{k}$ is lower for Disbelievers than for Believers. Indicating that Disbelievers act in much more organized manner over the 100 weeks than Beleivers.  

Climate change Believers, on the other hand, were not generally hostile toward Disbelievers, as the time series for climate change Believers tends to fluctuate over and under the threshold of 0. This indicates that climate change Believers communicate with in-group and out-group members with relatively similar emotional valence. However, on certain weeks, climate change Believers did also feature exceptionally high hostility scores. This suggests that climate change Believers may also behave in a hostile manner toward climate change Disbelievers, even if not over the long term.

\begin{figure}[ht]
    \includegraphics[width=0.95\linewidth]{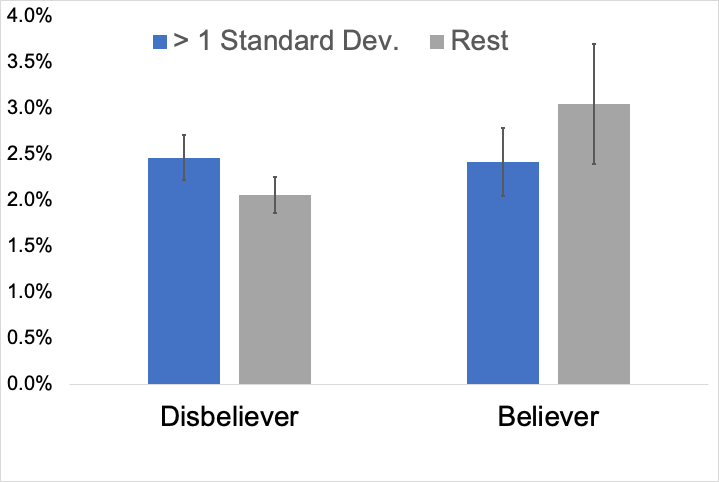}
    \caption{Percentage of the top 100 most frequent hashtags containing natural disaster-related words. The figure shows the percentage when the affective polarization metric is greater than 1 standard deviation or otherwise. The error bars represent ±1 standard errors. }
    \label{fig:barchart}
\end{figure}

To investigate instances where hostility between Believers and Disbelievers is high we compare those weeks with weeks where hostility is low. We define hostile weeks as those data points where $l_{k}$ is more than mean plus 1 standard deviation, i.e. from \fref{fig:timeline}, all the weeks where for Believers $l_{k} > 0.080 $ and for Disbelievers $l_{k} > 0.140$. The number of such weeks for Disbelievers where $l_{k} > 0.140 $ is 20 and for Believers where $l_{k} > 0.080 $ is 12. We look further into these weeks as examples of exceptional hostilie weeks.

\begin{figure}[ht]
    \includegraphics[width=0.95\linewidth]{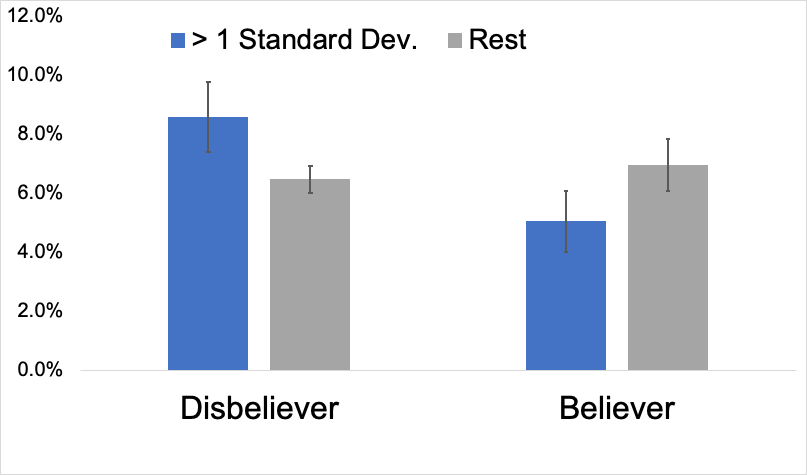}
    \caption{Percentage of tweets with at least one natural disaster-related word. The figure shows the percentage when the affective polarization metric is greater than 1 standard deviation or otherwise. The error bars represent ±1 standard errors.}
    \label{fig:barchart_word}
\end{figure}

Next, we use natural disaster-related words as a proxy to determine how natural disasters play a role in hostility between the two groups.  In \fref{fig:barchart} we look at the top 100 most frequent hashtags used within those groups to find the percentage of hashtags related to natural disasters. As expected, Believers use more natural disaster-related hashtags than Disbelievers. However, during the exceptional hostile weeks Believers use less of these hashtags. Interestingly, Disbelievers show the exact opposite behavior. Disbelievers use more natural disaster-related hashtags when they are more hostile towards Believers. We provide further evidence of this finding in \fref{fig:barchart_word}. In \fref{fig:barchart_word}, we look at the percentage of Tweets with at least one natural disaster-related word. We find similar patterns as mentioned above. Moreover, we find that a greater percentage of Tweets from Disbelievers mention natural disaster-related words compared to Believers. This indicates that Disbelievers are calling out natural disasters more when they are exceptionally hostile towards Believers compared to other weeks. 

\section{Discussion and Future Work}

Taken together, our findings suggest the importance of considering affective polarization in online discourse, particularly concerning the subject of climate change. Whereas past studies had shed light on the echo chamber dynamics which characterized intergroup communication surrounding climate change \cite{williams2015network}, we show how this polarization extends also to the realm of emotion in the form of affective polarization. We extend existing studies which highlight the role of incivility and personalized framing in encounters between climate change Believers and Disbelievers \cite{anderson2017social,van2020online} by introducing a scalable technique for analyzing relative intra- and intergroup interaction valence. This allowed us to quantify the extent of hostile communications between the two groups over a large-scale, long-term dataset - thereby validating existing findings in a generalizable manner as well as showing their relative stability over time.

Furthermore, we highlight the value of viewing polarization from an asymmetrical perspective. Related scholarship in political psychology underscores how ideological asymmetries underpin conflict dynamics across a variety of social issues \cite{jost2017ideological}. In other words, the participation of two groups within polarized discourse does not necessarily mean that both groups engage in conflict in the same way. Prior work illustrates that these findings translate robustly to the digital sphere - political elites or opinion leaders who share moralized content behave in distinct ways depending on their ideological orientations \cite{brady2017emotion}. The present work contributes to the literature by showing how these dynamics unfold the standpoint of the public at large concerning online climate change discourse.

Indeed, higher levels of hostility from Disbelievers present a specifically notable finding for social scientific scholarship on climate change discourse. Longitudinal analysis in prior work suggests that generalized climate change beliefs over time are increasing \cite{hamilton2015tracking,milfont2017public}, and climate change Disbelievers in particular are more susceptible to potential belief change \cite{jenkins2020partisan}. But significant cognitive barriers remain for fuller acceptance of anthropogenic causes for climate change and the corresponding urgency for responsive policy changes \cite{ballew2020does,carmichael2017elite}. Higher levels of hostility among climate change Disbelievers toward climate change Believers constitutes one such obstacle for further dialogue between the two groups. As past studies suggest, one psychological factor which impedes climate change Beliefs is not related to the climate at all, but anchors primarily on the feelings of dislike felt by one group towards the other \cite{van2018psychological}. Such challenges may thus persist in the form of further entrenchment of Disbelievers within interactional siloes and disengagement from intergroup communication altogether \cite{williams2015network}. Or as emergent studies show, they can also trigger what have been called `trench warfare dynamics' \cite{karlsen2017echo} - whereby Disbelievers persistently communicate with Believers but solidify their own cognitive immovability in the process. 

These insights are especially important to consider given our secondary set of findings. Our analysis suggests that further asymmetries arise between Believers and Disbelievers engagement with disaster words in relation to their levels of affective polarization. Although comparable levels are seen when both groups are within average levels of our metric, moments of increased affective polarization correlate with opposite behaviors for Believers and Disbelievers. Believers appear to shift to other areas of contention, such that their aggression is characterized by non-disaster topics. In contrast, Disbelievers' increased invocation of disaster terms points to more aggressive discussion of these catastrophes, albeit positioned in resistance to explanations related to anthropogenic climate change. This introduces another layer of intractable conflict in beliefs, as major climate events do not appear to invite susceptibility of belief change for Disbelievers. Instead, they potentially incite more vigorous psychological resistance.

Collectively, these findings point to significant benefits to studying affective polarization in online climate change discourse. Although social media discourse does not necessarily constitute a representative sample of a particular global population \cite{morstatter2013sample}, digital platforms like Twitter nonetheless constitute a vital space for public conversations about important issues like climate change. Hence, these findings paint a useful picture of public discourse as situated specifically in cyberspace, which may also bear implications for how digitally mediated science communication and public policy may also be designed and implemented \cite{bolsen2018us,kahan2015geoengineering}.

Besides the issue of demographic representativeness for online data, other limitations attend the present analysis. First, although we have a large number of tweets to characterize general affective behavior, however, it does not encompass those interactions which do not include our collection keywords. Second, the task of getting an aspect-level sentiment of each tweet towards other entities is a non-trivial task. We use Netmapper which has been used with reasonable accuracy for multiple sentiment level tasks \cite{carley2017trident,netmapper2}. The focus of this paper is on designing a framework to get affective polarization score between two competing groups and we do not make an effort to improve aspect-level sentiment scores. Last, in our analysis we use a list of natural disaster related words. Communication about the natural disasters could also happen using specific names related to these disasters, for example using ``Dorian'' instead of ``Hurricane Dorian''. Such analysis would require a more comprehensive list of natural disasters occurring around the world during the 100 weeks. This is out of scope for the current work.

Recognizing the foregoing limitations, we also consider avenues for future work in this area. On a conceptual level, researchers may wish to expand the binary system of climate change beliefs assumed here. Affectively polarized dynamics between multiple groups may be a more challenging yet also potentially informative line of inquiry to explore given the diversity of positions held with respect to this complex issue. Methodologically, computational analysis may extend our findings by performing more fine-grained characterization of the types of hostility expressed by both groups. Natural language processing (e.g., topic models) may offer one way forward in this regard. Acknowledging the non-neutrality of cyberspace, it would also be important to consider whether disinformation maneuvers may also be involved in shaping the wider climate change discussion. Inauthentic bot-like accounts and trolls may unduly influence different groups by manipulating the flow of information or amplifying intergroup aggressions; such factors have been seen in relation to other contentious issues and may potentially be present here as well \cite{carley2018social,broniatowski2018weaponized,bessi2016social}. 

Finally, taking flight from the digital scope of our research, further studies may fruitfully examine several hypotheses opened up by our results. For instance, social scientists may investigate actual levels of experienced hostility by climate change Believers and Disbelievers toward opposed groups. These evidence bases would be valuable to accumulate in cross-cultural settings, as well as over time - especially in connection with concurrent political shifts and natural climate-related developments like anomalous weather patterns and wider-ranging disasters \cite{carmichael2017elite,hamilton2015tracking,milfont2017public}.

\bibliographystyle{IEEEtran}
\bibliography{references}

\section{Appendix}
\label{sec:appendix}
\small{ List of natural disaster related words used in the analysis: avalanche,
blizzard,
bushfire,
cataclysm,
cloud,
cumulonimbus,
cyclone,
disaster,
drought,
duststorm,
earthquake,
erosion,
fire,
flood,
forestfire,
gale,
gust,
hail,
hailstorm,
heatwave,
high-pressure,
hurricane,
lava,
lightning,
low-pressure,
magma,
naturaldisasters,
nimbus,
permafrost,
rainstorm,
sandstorm,
seismic,
snowstorm,
storm,
thunderstorm,
tornado,
tremor,
tsunami,
twister,
violentstorm,
volcano,
whirlpool
whirlwind,
windstorm}

\end{document}